\documentclass[aps,twocolumn,groupeaddress,prl]{revtex4}
\usepackage{amssymb}
\usepackage{amsmath}
\usepackage{graphicx}
\usepackage{ulem}

\def\ttgq{\times 2e^2/h}
\def\tgq{2e^2/h}
\def\Isd{I}
\def\vsd{V_{\mathrm{sd}}}

\def\vgone{V_{\mathrm{g1}}}
\def\vgtwo{V_{\mathrm{g2}}}

\def\bpar{B_{\parallel}}
\def\He3{^3\mathrm{He}}
\def\pt7{$0.7\times2e^2/h$}

\def\tauns{\tau_{n,\sigma}}
\def\kb{k_{B}}
\def\te{T_{e}}
\def\Sipart{S_{I}^{\mathrm{P}}}
\def\gavg{g_{\mathrm{avg}}}
\def\NF{\mathcal{N}}
\def\n{\rho_{n}}
\def\gx{G_\mathrm{X}}
\def\Reff{R_{\mathrm{eff}}}
\def\fs{f_{\mathrm{s}}}
\def\fd{f_{\mathrm{d}}}

\begin{document}
\title{Shot-Noise Signatures of 0.7 Structure and Spin in a Quantum Point
Contact}
\author{L.\ DiCarlo\footnote[1]{These authors contributed equally to this work.}, Y.\ Zhang\footnotemark[1], D.\ T.\ McClure\footnotemark[1], D.\ J.\ Reilly, C.\ M.\ Marcus}
\affiliation{Department of Physics, Harvard University, Cambridge,
Massachusetts 02138, USA}
\author{L.~N.~Pfeiffer, K.~W.~West}
\affiliation{Bell Laboratories, Lucent Technologies, Murray Hill, NJ
07974, USA}
\date{\today}

\begin{abstract}
We report simultaneous measurement of shot noise and dc transport in
a quantum point contact as a function of source-drain bias, gate
voltage, and in-plane magnetic field. 
Shot noise at zero field exhibits an asymmetry related to the $0.7$ structure 
in conductance. The asymmetry in noise evolves smoothly into the symmetric signature
of spin-resolved electron transmission at high field. 
Comparison to a phenomenological model with 
density-dependent level splitting yields good
quantitative agreement.
\end{abstract}
\maketitle

Shot noise, the temporal fluctuation of current resulting from the
quantization of charge, is sensitive to quantum statistics,
scattering and many-body effects~\cite{Blanter00,Martin05}. Pioneering
measurements~\cite{Reznikov95, Kumar96,Liu98} of shot noise in
quantum point contacts (QPCs) observed the
predicted~\cite{QPCnoisetheory} suppression of shot noise below the
Poisson value due to Fermi statistics. In regimes where many-body
effects are strong, shot noise measurements have been exploited
to directly observe quasiparticle charge in strongly correlated
systems~\cite{de-Picciotto97,Saminadayar97,Jehl00} as well as to
study coupled localized states in mesoscopic tunnel
junctions~\cite{Safonov03} and cotunneling in nanotube-based quantum
dots~\cite{Onac06}.

Paralleling these developments, a large literature has emerged
concerning the surprising appearance of an additional plateau in
transport through a QPC at zero magnetic field, termed 0.7
structure. Experiment~\cite{pointsevenexpt, Oliver02,Roche04} and
theory~\cite{pointseventheory, Reilly05} suggest that 0.7 structure
is a many-body spin effect. Its underlying microscopic origin,
however, remains an outstanding problem in mesoscopic physics.  This
persistently unresolved issue is remarkable given the simplicity of
the device.

In this Letter, 
we report simultaneous measurements of the shot
noise at 2~MHz and dc transport in a QPC, exploring the noise
signature of the 0.7 structure and its evolution with in-plane
magnetic field $\bpar$. A suppression of the noise relative to that
predicted by theory for spin-degenerate transport~\cite{QPCnoisetheory} is
observed near \pt7 at $\bpar=0$, in agreement with
results from Roche {\it et al.}~\cite{Roche04} obtained at kHz
frequencies. This suppression evolves smoothly with increasing
$\bpar$ into the signature of spin-resolved
transmission. We find quantitative agreement 
between noise data and a  phenomenological model
for a density-dependent level splitting~\cite{Reilly05}, 
with model parameters extracted solely from conductance.

\begin{figure}[t]
\center \label{fig1}
\includegraphics[width=2.9in]{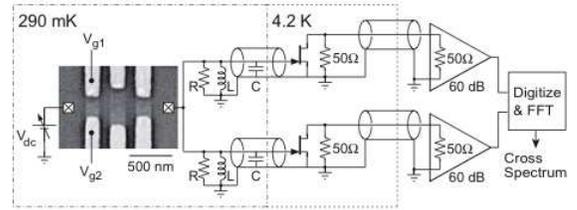}
\caption{\footnotesize{ Equivalent circuit near $2~\mathrm{MHz}$ of
the noise detection system measuring QPC noise by cross-correlation
on two amplification channels \cite{Techniques06}. The
scanning electron micrograph shows a
device of identical design to the one measured. The
QPC is formed by negative voltages $\vgone$ and $\vgtwo$ applied on
two facing electrostatic gates. All other gates on the device are
grounded.}}
\end{figure}

Measurements are performed on a gate-defined QPC fabricated on the
surface of a $\mathrm{GaAs}/\mathrm{Al}_{0.3}\mathrm{Ga}_{0.7}\mathrm{As}$
heterostructure grown by molecular beam epitaxy (see micrograph in Fig.~1). 
The two-dimensional electron gas $190~\mathrm{nm}$ below the surface has
a density of $1.7\times 10^{-11}~\mathrm{cm}^{-2}$ and mobility
$5.6\times10^6~ \mathrm{cm}^2/\mathrm{Vs}$. 
All data reported here were taken at 290~mK, the base temperature of a $\He3$
cryostat.

The differential conductance $g=dI/d\vsd$ (where $I$ is the current and $\vsd$ is the 
source-drain bias) is measured by lock-in technique with an applied
$25~\mu V_{\mathrm{rms}}$ excitation at 430~Hz~\cite{Techniques06}. The 
resistance $R_{s}$ in series with 
the QPC is subtracted at every applied $\bpar$  (see Fig.~2(a))~\cite{Bperpnote}. 

\begin{figure}[t]
\center \label{fig:cond}
\includegraphics[width=3.25in]{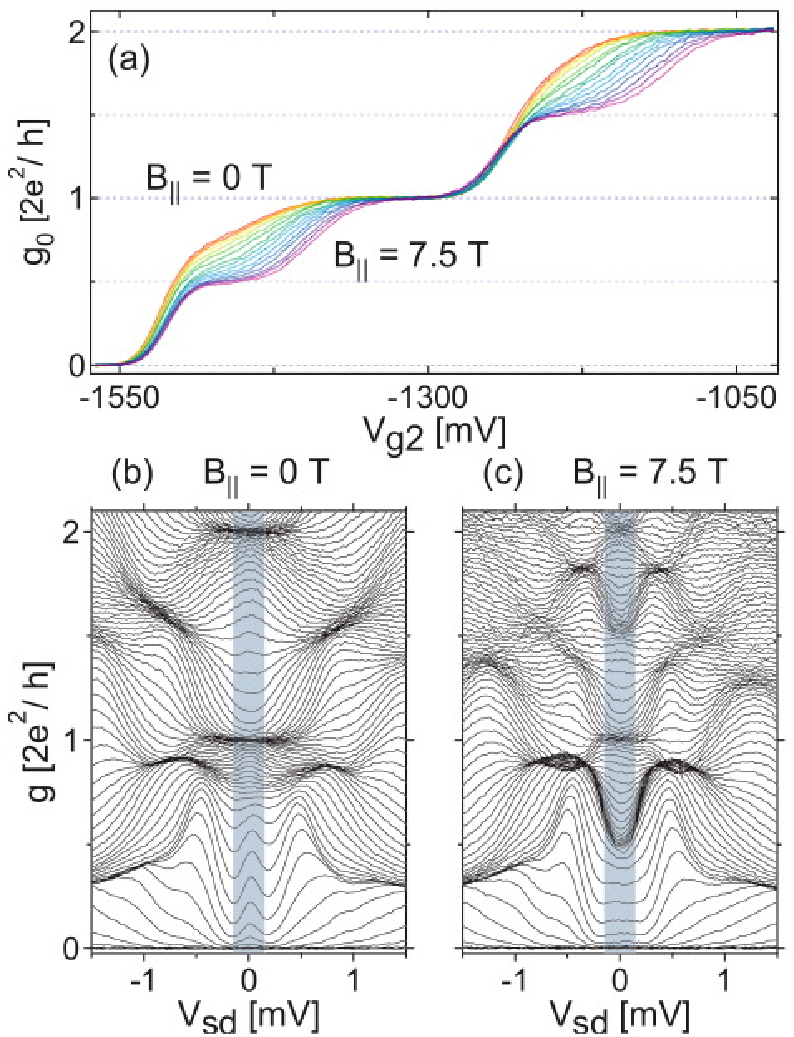}
\caption{\footnotesize{(color) (a) Linear conductance $g_0$ as a function
of $V_{\mathrm{g2}}$ ($\vgone = -\mathrm{3.2~V}$),  for $\bpar$
ranging from $0$ (red) to $7.5~\mathrm{T}$ (purple) in steps of
$0.5~\mathrm{T}$. The series resistance $R_s$ ranging from
430~$\mathrm{\Omega}$ at $\bpar=0$ to 730~$\mathrm{\Omega}$ at
$\bpar=7.5~\mathrm{T}$ has been subtracted to align the plateaus
at multiples of $\tgq$. (b,c) Nonlinear differential conductance 
$g$ as a function of $\vsd$, at $\bpar= 0$ (b) and 7.5~T (c), with
$V_{\mathrm{g2}}$ intervals of 7.5 and 5~mV, respectively. Shaded
regions indicate the bias range used for the noise measurements
presented in Figs.~3(b)~and~4.}}
\end{figure}

The QPC is first characterized at zero and finite $\bpar$ using dc conductance
measurements. Figure~2(a) shows linear-response conductance $g_0
=g(\vsd\sim 0)$ as a function of gate voltage $\vgtwo$, for $\bpar=0$
 to 7.5~T in steps of 0.5~T. The QPC shows the
characteristic quantization of conductance in units of $\tgq$ at
$\bpar=0$, and the appearance of spin-resolved plateaus at multiples
of $0.5\ttgq$ at $\bpar=7.5~\mathrm{T}$. Additionally, at $\bpar=0$,
a shoulder-like 0.7 structure is evident, which evolves smoothly into
the $0.5\ttgq$ spin-resolved plateau at high $\bpar$.

Figures~2(b) and 2(c) show $g$ as a function of $\vsd$ for evenly spaced 
$\vgtwo$ settings at $\bpar=0$ and 7.5~T, respectively.  In this representation,
linear-response plateaus in Fig.~2(a) appear as accumulated traces
around $\vsd=0$ at multiples of $\tgq$ for $\bpar=0$,  and at
multiples of $0.5\ttgq$ for $\bpar=7.5~\mathrm{T}$. At finite
$\vsd$, additional plateaus occur when a sub-band edge lies between the source 
and drain chemical potentials~\cite{Kouwenhoven89}. The features near $0.8\ttgq$ 
($\vsd \approx \pm750~\mu$V) at $\bpar=0$ cannot be explained in the context
of a single-particle picture~\cite{pointseventheory,pointsevenexpt}. These features are
related to the 0.7 structure around $\vsd=0$ and resemble the
spin-resolved finite bias plateaus at $\sim0.8\ttgq$ for
$\bpar=7.5$~T~\cite{pointsevenexpt}.

Turning now to noise measurements,
we consider the QPC noise in excess of thermal noise $4\kb\te g(\vsd)$. When
1/f and telegraph noise as
well as bias dependent heating are negligible (as shown to be the
case in these data) the excess noise is dominated by noise arising from the
partitioning of electrons
at the QPC, which we denote as partition noise, $\Sipart(\vsd)=S_{I}(\vsd)-4\kb\te
g(\vsd)$, where $S_{I}$ is the total QPC current noise spectral density. 
Note that $\Sipart$ is noise in excess of $4\kb\te g(\vsd)$ 
rather than $4\kb\te g(0)$ as considered in Refs.~\cite{Reznikov95,Roche04}. 

We measure $\Sipart$ near 2~MHz using the cross-correlation technique
shown schematically in Fig.~1 to suppress amplifier voltage noise~\cite{Techniques06,Kumar96}.
Two parallel channels amplify the voltage fluctuations across a
resistor-inductor-capacitor resonator that performs current-to-voltage conversion.
Each channel consists of a transconductance stage using a
high electron mobility transistor (HEMT) cooled  to 4.2~K, followed by  $50~\Omega$
amplification at room temperature. The amplified noise signals from both channels are sampled 
simultaneously by a 
digitizer, and their cross-spectral density calculated by fast-Fourier-transform.

The cross-spectral density is maximal at resonance, with a value
\begin{equation}
X_R^0=\gx^{2}\left(\Sipart \left(\frac{\Reff}{1+g R_s}\right)^2 + 4 \kb \te \Reff\right),
\end{equation}
where  $\gx$ is the geometric mean of the voltage gain of the amplification
channels,  $\te$ is the electron temperature and $\Reff$ is the 
effective resistance (at 2~MHz) between the HEMT gates and ground.
$\Reff$ is measured from the half-power bandwidth of the cross-spectral
density~\cite{Techniques06}. $\Sipart$ is extracted from simultaneous 
measurements of $X_{R}^{0}$, $g$ and $\Reff$
following calibration of $\gx$ and $\te$ using thermal noise.
At $\vsd=0$, where $\Sipart$  vanishes, 
$X_{R}^{0}= \gx^{2}\cdot4\kb\te\Reff$. At
elevated temperatures (3 to 5 K), where electrons are
well thermalized to a calibrated thermometer, a measurement of $X_{R}^{0}$
as a function of  $\Reff$ (tuned through $\vgtwo$) 
allows a calibration of $\gx=790~\mathrm{V/V}$. This gain is then used to 
determine from similar measurements 
the base electron temperature $\te=290~\mathrm{mK}$.

Figure~3 shows $\Sipart(\vsd)$ at $\bpar=0$ and fixed $\vgtwo$ for $\vsd$
between $-150~\mu\mathrm{V}$ and $+150~\mu\mathrm{V}$ 
(blue regions in Figs.~2(b) and 2(c)).  With an
integration time of  60~s at each bias point, the resolution in
$\Sipart$ is $1.4\times10^{-29}~\mathrm{A}^2/\mathrm{Hz}$,
equivalent to full shot noise $2eI$ of $I\sim 40~\mathrm{pA}$.
Open markers superimposed on the linear conductance trace in
Fig.~3(a) indicate $\vgtwo$ settings for which corresponding noise
data are shown in Fig.~3(b). $\Sipart$ vanishes with the QPC pinched off ($g(\vsd)=0$),
or on linear conductance plateaus, which shows that bias-dependent electron heating is 
not significant~\cite{Kumar96}. In contrast, for $g \approx$ 0.5 and $1.5\ttgq$,
$\Sipart$ grows with $|\vsd|$ and shows a transition from quadratic to linear dependence
~\cite{Reznikov95,Kumar96,Liu98}, demonstrating the absence of noise from resistance fluctuations.

Solid curves superimposed on the $\Sipart(\vsd)$ data in Fig.~3(b) 
are fits to the form
\begin{equation}
\Sipart(\vsd) = 2\frac{2e^2}{h}\NF\left[e
\vsd\coth\left(\frac{e\vsd}{2\kb\te}\right)-2\kb\te\right],
\end{equation}
with the \textit{noise factor} $\NF$ as the only free fitting parameter. Note
that $\NF$ relates $\Sipart$ to $\vsd$, in contrast to the
Fano factor, which
relates $\Sipart$ to $I$ \cite{Blanter00,Martin05}. The form of this fitting function is 
motivated by mesoscopic scattering theory~\cite{QPCnoisetheory,Blanter00,Martin05}, 
where transport is described by transmission coefficients 
$\tauns$ ($n$ is the transverse mode index and $\mathrm{\sigma}$ denotes spin) 
and partition noise originates from the partial transmission of incident electrons.
Within scattering theory, the full expression for  $\Sipart$ is 
\begin{equation}
\Sipart(\vsd) =
\frac{2e^2}{h}\int\sum_{n,\sigma}\tau_{n,\sigma}(\varepsilon)(1-\tau_{n,\sigma}(\varepsilon))(\fs-\fd)^2
d\varepsilon,
\end{equation} 
where $f_{\mathrm{s(d)}}$ is the Fermi function in
the source (drain) lead. Eq.~(2) follows from Eq.~(3) only 
for the case of constant transmission across the energy window of transport, 
with  $\NF=\frac{1}{2}\sum\tauns(1-\tauns)$. For spin-degenerate
transmission, $\NF$ vanishes at multiples of $\tgq$ and reaches
the maximal value 0.25 at odd multiples of $0.5\ttgq$. 

We emphasize that while Eq.~(2) is motivated by scattering theory, 
the value of $\NF$ extracted from fitting with Eq.~(2) simply provides a 
way to quantify the $\Sipart(\vsd)$ for each $\vgtwo$.
We have chosen the bias range  $e |\vsd| \lesssim  5 \kb \te$ 
for fitting $\NF$ to minimize the effects of nonlinear transport while 
extending beyond the quadratic-to-linear
crossover in noise that occurs on the scale  $e\vsd\sim2\kb\te$.

\begin{figure}[t!]
\center \label{fig:vs}
\includegraphics[width=3.25in]{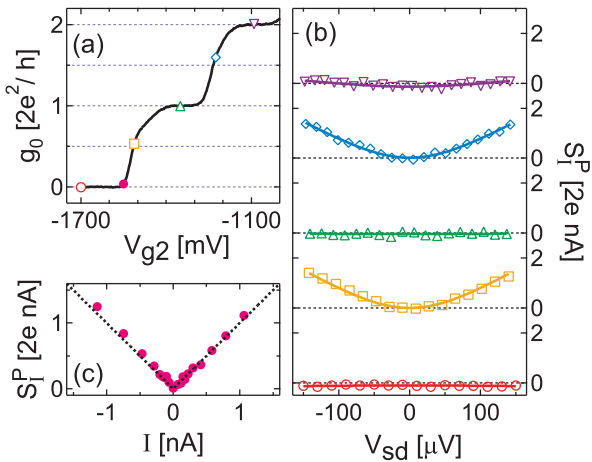}
\caption{\footnotesize{(color) (a) Linear conductance $g_0$ as a function
of $\vgtwo$ at $\bpar=0$. Markers indicate $\vgtwo$
settings for the noise measurements shown in (b) and (c). (b) Measured $\Sipart$ as a function of
$\vsd$, for conductances near 0 (red), 0.5 (orange), 1 (green), 1.5
(blue), and 2~$\ttgq$ (purple). Solid lines are best-fits to Eq.~(2)
using $\NF$ as the only fitting parameter. In order of increasing
conductance, best-fit $\NF$ values are 0.00, 0.20, 0.00, 0.19, and
0.03. (c) $\Sipart$ as a function of dc current $\Isd$ with the QPC
near pinch-off. The dotted line indicates full shot noise
$\Sipart=2e|I|$, comparable to results in Ref.~\cite{Chen06}.}}
\end{figure}

\begin{figure}[h!] \center \label{fig4}
\includegraphics[width=3.25in]{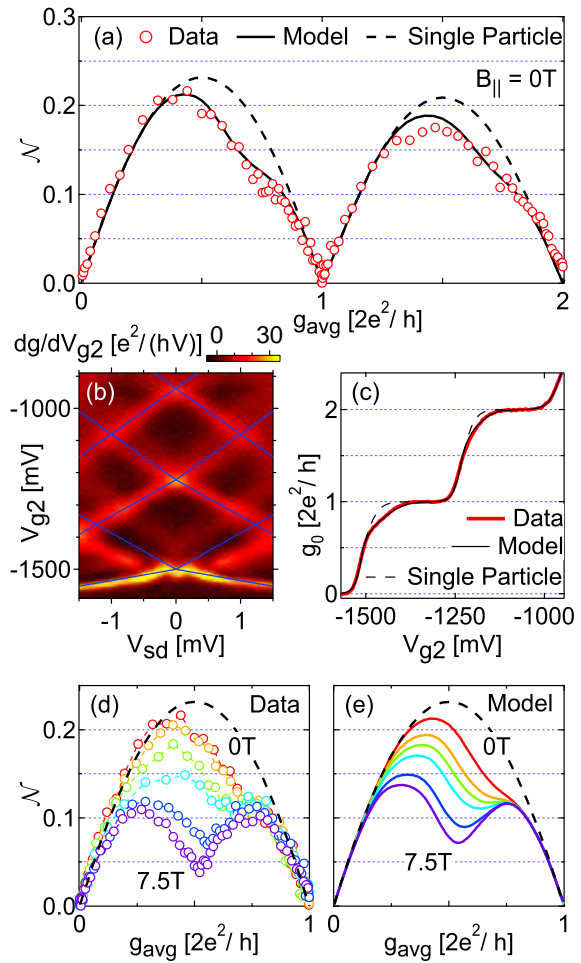}
\caption{\footnotesize{(color) (a) Experimental $\NF$ 
as a function of $g_{\mathrm{avg}}$ at $\bpar=0$ (red circles) along with
model curves for nonzero (solid) and zero (dashed) proportionality
of splitting, $\gamma_{n}$ (see text). (b) Transconductance $dg/d\vgtwo$ as a
function of bias voltage $\vsd$ and gate voltage $\vgtwo$.
Blue lines trace the alignment of sub-band edges
with source and drain chemical potentials; their slope and
intersection give the conversion from $\vgtwo$ to
energy and the energy spacing between modes, respectively
~\cite{Patel91,Detail02}. (c) Measured linear conductance (red)
as a function of $\vgtwo$ at $\bpar = 0$, and linear conductance
calculated with the model (black solid) with best-fit values for $\omega_{x}$
and $\gamma_{n}$. Single-particle conductance model takes $\gamma_{n} = 0$
(black dashed). (d) Experimental $\NF$ as a
function of $g_{\mathrm{avg}}$ in the range $0-1\ttgq$,
at $\bpar = $ 0~T (red), 2~T (orange), 3~T (green), 4~T (cyan), 6~T (blue),
and 7.5~T (purple). (e) Model curves for $\NF(\gavg)$ (see text). Dashed curves
in (d) and (e)
show the single-particle model ($\gamma_{n} = 0$) at zero field for comparison.
}}
\end{figure}

The dependence of noise factor on QPC conductance at $\bpar=0$ is shown in
Fig.~4(a), where $\NF$ is extracted from measured $\Sipart(\vsd)$ at 90
values of $\vgtwo$. The horizontal axis, $\gavg$, is the
average of the differential
conductance over the bias points where noise was measured.
 $\NF$ has the shape of a dome,
reaching a maximum near odd multiples of $0.5\ttgq$ and vanishing at
multiples of $\tgq$. The observed $\NF (\gavg)$ 
deviates from the spin-degenerate, energy-independent scattering theory in two
ways. First, there is a reduction in the maximum amplitude of $\NF$ below
$0.25$. Second, there is an asymmetry in $\NF$ with
respect to $0.5\ttgq$, resulting from a noise reduction near the
0.7 feature. A similar but weaker asymmetry is observed 
about $1.5\ttgq$.

The dependence of $\NF (\gavg)$ on $\bpar$ is shown in Fig.~4(d).
 $\NF$ is seen to evolve smoothly from a single
asymmetric dome at $\bpar=0$ to a symmetric double-dome at
$7.5~\mathrm{T}$, the latter a signature of spin-resolved electron
transmission.
Notably, near $0.7\ttgq$,  $\NF$ appears insensitive  to $\bpar$, in 
contrast to the dependence of $\NF$ near $0.3\ttgq$.

We find that all features in noise data are well accounted for within a simple
phenomenological model in which the twofold degeneracy of QPC
mode $n$ is lifted by a splitting $\Delta
\varepsilon_{n,\sigma}= \sigma\cdot\n\cdot\gamma_n$, that grows linearly with
1D density $\n$ (with proportionality $\gamma_n$) within that mode. Here, $\sigma =
\pm
1/2$ and $\n \propto \sum_{\sigma}\sqrt{\mu-\varepsilon_{n,\sigma}}$, ($\mu$ 
is the chemical potential).
The lever arm converting 
from $\vgtwo$ to energy (and hence $\n$) as well transverse mode spacing are
extracted from transconductance $(dg/dV_{g2})$ data (Fig.~4(b)).  
Assuming an energy-dependent transmission, $\tauns(\varepsilon) = 1/(1+e^{2\pi(\varepsilon_{n,\sigma}-\varepsilon)/\hbar\omega_{x}})$, 
appropriate for a saddle-point potential with curvature parallel to the
current described by $\omega_{x}$ \cite{Buttiker90}, the value for $\omega_{x}$
is found by fitting linear conductance below $0.5\ttgq$ (below $1.5\ttgq$
for the second mode), and $\gamma_n$ is obtained from a fit to 
conductance above $0.5 (1.5)\ttgq$, where (within the model)
the splitting is largest (see Fig.~4(c)). For the QPC studied, we find $\hbar\omega_x$ is $\sim500(300)~\mu\mathrm{eV}$ and
$\gamma_{1(2)}\sim0.012(0.008)~e^2/4\pi\epsilon_0$ for the first
(second) transverse modes. Note that the splitting is two orders of
magnitude smaller than the direct Coulomb energy of electrons spaced by $1/\n$.

Using these parameters, model values for $\Sipart(\vsd)$ are then calculated
using the full Eq.~(3), and $\NF$ is extracted by fitting the model  $\Sipart(\vsd)$
to  Eq.~(2).  The resulting model values of $\NF(\gavg)$ at $\bpar=0$ are shown
along with the experimental data in Fig.~4(a). Also shown for comparison are the
model values only accounting for energy dependent transmission but no
splitting ($\gamma_{n}=0$). The overall reduction of $\NF$ 
arises from a variation in transmission across the 
$150~\mu\mathrm{V}$ bias window, which is comparable to
$\hbar\omega_x$. Asymmetry of the model values for $\NF$ about $0.5$ and $1.5\ttgq$
require nonzero $\gamma_{n}$.

We include magnetic field in the model with corresponding simplicity
by assuming a g-factor of 0.44 and adding the Zeeman splitting to the
density-dependent splitting~\cite{Detail03} maintaining the parameters
obtained above. The resulting model values for $\NF$ are shown in Fig.~4(e), next to the
corresponding experimental data (Fig.~4(d)).
Experimental and model values for $\NF$ show comparable evolution in $\bpar$:
the asymmetric dome at $\bpar=0$ evolves smoothly into a  double dome at 7.5~T, and
for conductance $\gtrsim0.7\ttgq$,  the curves  for all magnetic
fields overlap closely. Some differences are observed between
data and model, particularly for  $\bpar=7.5~\mathrm{T}$.
While the experimental double-dome is symmetric with respect to the
minimum at $0.5\ttgq$, the theory curve remains slightly asymmetric with a
less pronounced minimum. We find that setting the g-factor to
$\sim0.6$ in the model reproduces the measured symmetrical
double-dome as well as the minimum value of $\NF$ at $0.5\ttgq$.
This observation is consistent with previous reports of an enhanced
g-factor in a QPC at low-density~\cite{pointsevenexpt}.

We thank H.-A.~Engel,  M.~Heiblum, L.~Levitov and A.~Yacoby for valuable
discussions, and S.~K.~Slater, E.~Onitskansky and N.~J.~Craig for
device fabrication. We acknowledge support from NSF-NSEC, 
ARO/ARDA/DTO and Harvard University.


\begin{thebibliography}{39}
\bibitem[{\citenamefont{Bperpnote}(05)}]{Bperpnote}
A 125~mT perpendicular magnetic field is applied throughout to minimize
bias-dependent electron heating~\cite{Kumar96}.
Perpendicular fields of this size do not affect the main results presented here.
\bibitem[{\citenamefont{Detail02}(05)}]{Detail02}
Two crossing points are observed at finite bias between the first and second modes. 
The model attributes this to spin splitting in the first mode.  
The midpoint is taken as the crossing point for the blue lines.
\end{thebibliography}
\end{document}